# Smartwatch-derived Acoustic Markers for Deficits in Cognitively Relevant Everyday Functioning

Yasunori Yamada[1*], Kaoru Shinkawa[1], Masatomo Kobayashi[1], Miyuki Nemoto[2], Miho Ota[2], Kiyotaka Nemoto[2], Tetsuaki Arai[2]

[1]Digital Health, IBM Research, Tokyo, Japan

[2]Department of Psychiatry, Division of Clinical Medicine, Institute of Medicine, University of Tsukuba, Ibaraki, Japan

*ysnr@jp.ibm.com

*Abstract*—Detection of subtle deficits in everyday functioning due to cognitive impairment is important for early detection of neurodegenerative diseases, particularly Alzheimer's disease. However, current standards for assessment of everyday functioning are based on qualitative, subjective ratings. Speech has been shown to provide good objective markers for cognitive impairments, but the association with cognition-relevant everyday functioning remains uninvestigated. In this study, we demonstrate the feasibility of using a smartwatch-based application to collect acoustic features as objective markers for detecting deficits in everyday functioning. We collected voice data during the performance of cognitive tasks and daily conversation, as possible application scenarios, from 54 older adults, along with a measure of everyday functioning. Machine learning models using acoustic features could detect individuals with deficits in everyday functioning with up to 77.8% accuracy, which was higher than the 68.5% accuracy with standard neuropsychological tests. We also identified common acoustic features for robustly discriminating deficits in everyday functioning across both types of voice data (cognitive tasks and daily conversation). Our results suggest that common acoustic features extracted from different types of voice data can be used as markers for deficits in everyday functioning.

## I. Introduction

With the aging of populations worldwide, the impact of cognitive impairment on everyday functioning has become a serious health and social problem. Deficits in everyday functioning in older adults can be associated with reduced quality of life for both themselves and their caregivers, increased economic burden, and inability to live independently [1–4]. Additionally, increasing evidence demonstrates that subtle changes in everyday functioning are observed early in neurodegenerative diseases of aging [5–7]. In particular, deficits in everyday functioning due to cognitive impairment constitute a core characteristic of dementia [8,9]. Subtle changes in cognitively complex everyday activities, such as cooking, shopping, and managing paperwork, have been shown to be present in the prodromal stage of dementia (i.e., mild cognitive impairment) [10,11] and even in the preclinical stage of Alzheimer's disease [12,13]. Such changes can also have prognostic value for predicting future incidents of mild cognitive impairment or dementia [14,15], as well as a faster rate of subsequent disease progression [16,17]. Therefore, early detection of deficits in everyday functioning is critically important.

Although there is no gold-standard method for measuring subtle changes in everyday functioning, performance- and questionnaire-based measures have been widely used. Performance-based measures rate a subject's performance while performing multiple tasks simulating daily activities under the direct observation of a trained professional. The representative examples include the University of California, San Diego Performance-Based Skills Assessment, and the Harvard Automated Phone Task, which have shown promise in detecting changes across the spectrum of Alzheimer's disease [18,19]. The limitations of these measures are that they can be time consuming, and their artificial conditions can negatively impact their ecological validity and reliability [1,20]. On the other hand, questionnaire-based measures include both self-reported and informant-reported ratings of everyday functioning. The representative examples are the Everyday Cognition (ECog) scale [1] and the Amsterdam Instrumental Activities of Daily Living Questionnaire [19]. Although the relative sensitivity and predictive capability of self-reported versus informant-reported measures have yielded mixed results [17,22], both have proven beneficial for identifying individuals at risk for disease progression, particularly in early disease stages [22,23]. Although these measures are easy to administer and convenient, their potential limitation is that subjective ratings can be affected by multiple factors, including depressive mood, anxiety, and stress [24–26]. Hence, if subtle changes in everyday functioning could be objectively detected by an easy-to-perform method, it would help early detection of at-risk individuals.

Digital health technologies involving wearable sensors and mobile devices have been expected to offer novel, less burdensome tools for frequently and objectively monitoring health-relevant behavior changes in everyday life [24,26,27]. Among the devices for sensing, wearable ones such as smartwatches have particular advantages for passive data collection, which allows continuous data acquisition with higher adherence and without learning effects [26]. Examples of cognition-relevant markers derived from wearable sensors include accelerometer-based physical activity metrics [28] and electrocardiography-based metrics of heart rate variability [29]. In addition, although to the best of our knowledge no study has investigated voice-based markers for cognitive impairments derived from wearable sensors, smartwatches have gained attention as a tool for capturing in-the-wild voice data that would better reflect the user's health status [30].

In terms of behavioral modalities to produce digital markers, the human voice is a promising data source for detecting subtle cognitive impairments in older adults [26,31,32]. Numerous studies have investigated voice data collected during cognitive tasks, typically in

This work was supported by the Japan Society for the Promotion of Science, KAKENHI (grant 19H01084).

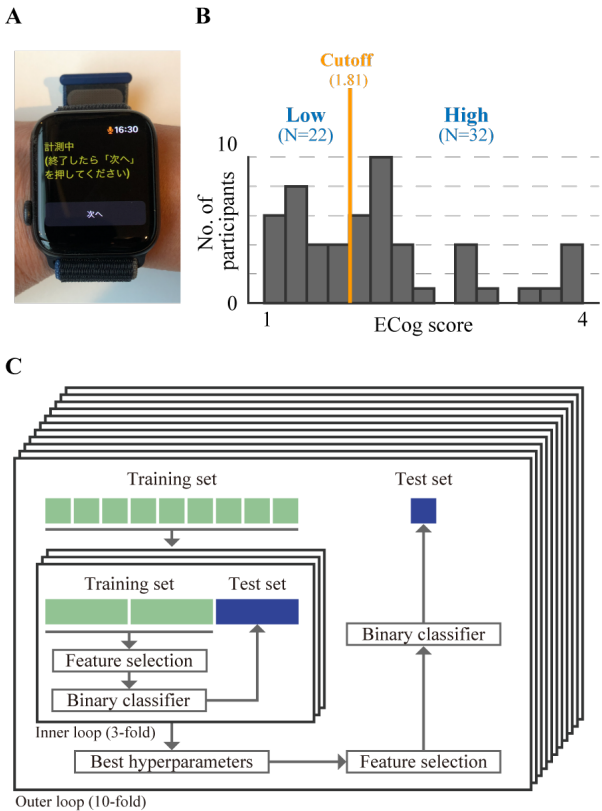

Fig. 1. Overview of the data analysis on voice responses to cognitive tasks and daily life questions, which were recorded with a smartwatch-based application for detecting individuals with deficits in everyday functioning. **A**. Smartwatch-based application for collecting voice data. **B**. Histogram of the study participants' Everyday Cognition (ECog) scores. **C**. 10×3 nested cross-validation procedure in a supervised machine learning model.

neuropsychological assessments, and they have reported speech and language disturbances due to cognitive impairments across the disease spectrum [33–36]. For instance, speech and language disturbances are evident even in early stages of dementia, including the prodromal and preclinical stages of Alzheimer's disease [37]. They are also predictors for disease progression from normal cognition to mild cognitive impairment and to Alzheimer's disease [33,34]. These disturbances include changes in linguistic features, such as reduced speech expressiveness and increased repetition [35,36,38], and changes in acoustic features, such as a reduced speech rate, increased pauses, and differences in spectrum-based features [33,35,36]. By leveraging these linguistic and acoustic features, machine learning models and deep learning models have succeeded in reliably detecting subtle cognitive impairments in the early stages of Alzheimer's disease [34,36,39]. Although no study has investigated whether voice data can be used for capturing deficits in cognitively relevant everyday functioning, it is reasonable to explore the feasibility of voice-based applications for evaluating everyday functioning in older adults. In addition, in terms of the types of voice data, recent studies have shown that cognitive impairments can elicit discernible differences even in data collected when an individual is not explicitly performing a specific cognitive task. Examples include interviews about non-clinical topics [40,41], storytelling [42,43], phone conversations [38,44], and conversations with a voice assistant, an AI chatbot and humanoid robot [45–47]. The capability to infer a subject's cognition-relevant status from various types of voice data would help extend the application scope and increase the accessibility.

In this study, we investigate the feasibility of a smartwatch-based application for automatically detecting deficits in everyday functioning by capturing subtle cognitive impairments from voice data. To this end, we collected voice data with an Apple Watch from 54 older adults performing cognitive tasks and daily conversation as possible application scenarios. We also administered the ECog scale as a measure of cognition-relevant everyday functioning. We then investigated whether machine learning classification models could use acoustic features derived from voice data to differentiate individuals with high and low ECog scores. In addition, we explored which acoustic features were robustly discriminative for deficits in everyday functioning across different types of voice data. Accordingly, we compared the acoustic features that most contributed to the classification models between two different settings (cognitive tasks and daily conversation) and identified common acoustic features.

Our contributions are three-fold. First, we developed a smartwatch-based application and collected voice data from older adults along with a measure of everyday functioning. Second, we demonstrate that deficits in everyday functioning can be detected from acoustic features in voice data collected during both cognitive tasks and daily conversation. Third, we suggest that common acoustic features extracted from different types of voice data can be used as markers for deficits in everyday functioning.

## II. METHODS

This study was conducted under the approval of the Ethics Committee, University of Tsukuba Hospital (H29-065). The participants were outpatients from the Department of Psychiatry, University of Tsukuba Hospital, Ibaraki, Japan. We intentionally included those with and without dementia to cover a wide spectrum of cognitive impairments. All participants provided written informed consent after the study procedures had been fully explained to them. All examinations were conducted in Japanese.

### A. Assessment of everyday functioning

Everyday functional limitations due to cognitive impairments were assessed with the ECog scale [1]. The ECog scale was designed to capture subtle functional changes that can be observed in early disease stages such as mild cognitive impairment and to measure functional abilities that can be linked to specific cognitive domains [1]. The scale comprises 39 items covering six cognitive domains: memory, language, visual perception, planning, organization, and divided attention. Each item is rated by comparing the participant's current level of everyday functioning to the level 10 years earlier. The response options encompass a four-point scale: 1 = better or no change, 2 = occasionally worse, 3 = consistently a little worse, 4 = consistently much worse. The ECog score is calculated as an average score for all completed items and ranges between 1 and 4. Thus, higher scores indicate greater limitations on everyday functioning. We used 1.81 as a cutoff score for discriminating between cognitively normal older adults and cognitively impaired older adults (i.e., those with either mild cognitive impairment or dementia) [48]. We used

TABLE I. DEMOGRAPHICS AND COGNITIVE/CLINICAL MEASURES OF THE PARTICIPANTS.

| Variable | All (N=54) | | Everyday Cognition (ECog) | | | | *p*-value |
|---|---|---|---|---|---|---|---|
| | | | Low (N=22) | | High (N=32) | | |
| Age, years | 76.1 | (6.0) | 76.3 | (5.9) | 76.1 | (6.1) | 0.902 |
| Sex, female, *n* (%) | 27 | (50.0) | 13 | (59.1) | 14 | (43.8) | 0.268 |
| Education, years | 13.2 | (2.7) | 13.1 | (2.6) | 13.2 | (2.8) | 0.964 |
| Geriatric Depression Scale [a] | 4.1 | (3.6) | 3.3 | (3.4) | 4.7 | (3.6) | 0.188 |
| Clock Drawing Test [b] | 8.9 | (2.2) | 9.1 | (1.8) | 8.7 | (2.4) | 0.519 |
| Logical Memory IA [c] | 7.2 | (5.2) | 8.6 | (4.2) | 6.2 | (5.7) | 0.106 |
| Logical Memory IIA [c] | 4.1 | (4.4) | 4.3 | (3.5) | 4.0 | (4.9) | 0.778 |
| Frontal Assessment Battery [d] | 12.3 | (2.5) | 12.2 | (2.2) | 12.4 | (2.8) | 0.778 |
| Trail Making Test part A [e] | 68.3 | (62.2) | 60.3 | (20.0) | 74.1 | (79.6) | 0.440 |
| Trail Making Test part B [e] | 148.7 | (98.4) | 142.2 | (108.3) | 153.8 | (89.8) | 0.698 |
| Mini-Mental State Examination [f] | 24.24 | (4.73) | 26.00 | (3.37) | 23.03 | (5.13) | **0.023** |
| Instrumented Activity of Daily Living [g] | 6.62 | (2.28) | 7.73 | (0.69) | 5.84 | (2.67) | **0.002** |
| Everyday Cognition (ECog) [h] | 2.06 | (0.80) | 1.37 | (0.24) | 2.54 | (0.69) | **<0.001** |

Most data are shown as means (with standard deviations in parentheses) and were assessed with *t*-test, while the sex data are shown as numbers (with percentages) and were assessed with a chi-square test. Bold values highlight statistically significant differences. [a]The total possible score ranges from 0 to 15. [b]The total possible score ranges from 0 to 10. [c]The total possible score ranges from 0 to 25. [d]The total possible score ranges from 0 to 18. [e]The total possible score ranges from 0 to 300. [f]The total possible score ranges from 0 to 30. [g]The total possible score ranges from 0 to 8. [h]The total possible score ranges from 1 to 4.

the self-reported version of the ECog and translated the original questionnaire into Japanese with the author's consent.

*B. Voice data collection*

To investigate two representative application scenarios, we collected voice data during cognitive tasks and daily conversation. The cognitive tasks were selected from those that are typically used in neuropsychological assessments for screening cognitive impairments and require spoken responses. Daily conversation was simulated by using daily life questions regarding, for example, a recent dinner menu and future travel plans [49]. The participants sat down and were asked to respond to questions in an audio recording. The participants' voice responses were recorded by an Apple Watch series 6 worn on the wrist of their non-dominant hand (44.1 kHz; 16 bits; Fig. 1A). The voice data collection was conducted in the same quiet room with low reverberation for all participants. An experimenter sat behind a partition to remain hidden from the participant's view.

For the cognitive tasks, five tasks were included: counting backward, subtraction, phonemic and semantic verbal fluency, and picture description. The counting-backward task entailed counting back from 305 to 290. The subtraction task required the participant to subtract 3 from 925 and continue subtracting 3 until the result was below 900. The phonemic and semantic verbal fluency tasks respectively entailed producing as many nouns beginning with the Japanese mora "ta" and as many animal names as possible in 60 seconds. The picture description task required the participant to describe everything seen in the Cookie Theft picture from the Boston Diagnostic Aphasia Examination [50]. Acoustic and linguistic features derived from these tasks have been shown to be discriminative between cognitively normal patients and patients with mild cognitive impairment and Alzheimer's disease [33–39,51].

For the simulated daily conversation, we used five daily life questions as follows. The first two questions were related to past experiences with a fun childhood activity and a recent dinner menu. For these questions, the participant was asked to explain a fun activity that the participant had done as a child and to tell what the participant had eaten for dinner yesterday and the day before yesterday. The next two questions were associated with future expectations regarding risk planning and future travel planning. For risk planning, the participant was asked to explain what actions they would take and in what order if an earthquake struck while they were at home. For travel planning, the participant was asked to choose the mountains or the sea as a future travel destination for the summer and give three reasons for the choice. The last question was related to general knowledge: the participant was asked to explain a Japanese traditional event, Shichi-Go-San, to someone who was unfamiliar with it. The actual sentences of these daily life questions have been described elsewhere [49]. Acoustic features derived from responses to these daily life questions have also been reported to show discernible differences between the conditions of cognitively normal and mild cognitive impairment [40,41,49].

*C. Participants and dataset*

A total of 54 older individuals completed the everyday functioning assessment and the voice data collection (20 dementia patients (37.0%); 27 female (50.0 %); 58–91 years; mean (SD) age, 76.2 (6.0) years; Table I). The mean score on the ECog scale was 2.06 (SD: 0.80; participant range, 1.00–3.95; possible range, 1–4; Fig. 1B). We divided the participants into two groups ("high" and "low") by the cutoff score. In our sample, 32 older adults (14 female, 18 male; 59.2% of the participants) scored equal to or greater than the cutoff score, which indicated more problems in cognition-relevant everyday functioning.

For the speech data collection, 44 of the 54 participants responded to all 10 questions. The remaining participants did not complete one or more questions because they either could not follow the instructions, produced no response related to a task, or verbally refused to perform a task. Specifically, seven participants did not respond to one question (two participants for the phonemic verbal fluency task, and five participants for either the counting-backward, subtraction, or picture description task, the dinner menu question, or the risk planning question). One participant did not respond to both the general knowledge and risk planning questions. One participant did not respond to three tasks (the semantic and phonemic verbal fluency tasks and the picture description task). One participant responded only to the semantic verbal fluency and picture description tasks and the risk and travel planning questions. The average total length of the voice responses to the five tasks was 238.1 s (SD: 73.8 s) for the cognitive tasks and 136.2 s (SD: 95.4 s) for the daily life questions. The average durations of the responses to each question varied, with a range of 34.2–63.0 s for the cognitive tasks and 24.3–30.1 s for the daily life questions.

For all participants, in addition to the everyday functioning assessment and the voice data collection, neuropsychologists conducted following neuropsychological and clinical assessments (Table I): Mini-Mental State Examination (MMSE) to assess global cognition, immediate and delayed recall of Logical Memory Story A from the Wechsler Memory Scale-Revised for episodic memory, the Frontal Assessment Battery for executive function, Trail Making Test part-A for information processing speed and Trail Making Test part-B for executive function and attention, the Clock Drawing Test for visuospatial function, and the Geriatric Depression Scale to assess depressive symptoms. The participants also answered Lawton Instrumental Activities of Daily Living (IADL) and self-reported instruments about their education level. These measures did not statistically differ between the ECog high and low groups (*t*-test, $p > 0.05$), except for MMSE ($p = 0.023$) and IADL ($p = 0.002$).

*D. Voice data analysis and feature extraction*

In this study, we aimed to explore acoustic markers that would enable robust detection of individuals with a greater degree of functional limitation regardless of the content of voice data, including word production and spontaneous speech. Thus, we did not include linguistic features and instead focused on acoustic features. Specifically, we extracted a total of 42 acoustic features from each voice response. These acoustic features were determined on the basis of previous studies on Alzheimer's disease and related neurological disorders [35,36,49,52]. In the analysis, we regarded the responses to each question as independent samples. The features were extracted from the non-silent segments of each sample, by eliminating silent segments before the calculation.

The acoustic features comprised four types related to the voice spectrum, pitch, formant, and voice quality. The spectrum-based features were based on mel-frequency cepstral coefficients (MFCCs). MFCCs have been used to characterize a human voice's frequency distribution at a specific time by accounting for the response properties of the human auditory system [53]. Statistical values for MFCCs and their derivatives have been shown to discriminate patients with Alzheimer's disease and mild cognitive impairment from cognitively healthy older adults [35,36]. In particular, reduced MFCC variations are thought to be associated with changes in patients' voices, which are expressed as monotony and dullness of speech in clinical descriptions [54]. Previous studies suggested that these features can exhibit discernible differences for cognitive impairments during both cognitive tasks and daily conversation [49,52,55]. We computed the first 12 MFCCs along with their first-order derivatives ($\Delta$) and second-order derivatives ($\Delta\Delta$). For MFCC extraction, we used a sliding window size of 1102 frames (25 ms) with a shift of 441 frames (10 ms). The $\Delta$MFCC and $\Delta\Delta$MFCC features were calculated by local estimation of the derivatives via Savitzky-Golay filtering [56]. We also used pitch variation and the first and second formants (F1 and F2) as acoustic features. Reduced pitch variation can be related to reduced inflection, while changes in formants can capture inadequate vowel formation associated with limited movements of the articulators and particularly of the tongue due to increased muscle tension, for example [57]. Although these features have frequently been used for capturing changes due to mental health conditions related to depression and suicidality [54] as well as perceived loneliness [58], studies have also shown their usefulness for detecting cognitive impairments [49]. Lastly, the voice quality features comprised the jitter, shimmer, and harmonics-to-noise (HNR) ratio. The jitter and shimmer have been used to measure pathological voice quality by measuring cycle-to-cycle variations of the fundamental frequency and loudness [59]. Increased jitter and shimmer, particularly during held vowels, have consistently been observed in the voices of patients with ailments including Alzheimer's disease [47,60] and other neurological disorders such as Parkinson's disease [61,62]. Although these changes in speech data other than held vowels have yielded mixed results, studies have reported discernible changes even in the word production and spontaneous speech of patients with cognitive impairments [63]. The jitter feature was calculated as the average absolute difference between the length of consecutive glottal periods, divided by the average length of the glottal period [59]. The shimmer feature was calculated as the average absolute difference between the amplitudes of consecutive periods, divided by the average amplitude. The HNR represents the degree of acoustic periodicity and is often used as a measure of hoarseness in an individual's voice. A decreased HNR has been reported for the voices of patients with Alzheimer's disease and Parkinson's disease [64,65]. This measure has also been reported to be negatively correlated with depression severity [64]. It was computed as the ratio of a periodic signal's energy to the energy of the noise in the signal [59].

In summary, we used the variances of the first 12 MFCCs, $\Delta$MFCCs, and $\Delta\Delta$MFCCs, and the pitch variation, F1, F2, shimmer, jitter, and HNR as acoustic features. These features were calculated by using the following Python (v3.8) audio processing libraries: librosa (v0.8.0) [67] and Signal_Analysis (v0.1.26) [68].

*E. Machine learning analysis*

We used supervised machine learning with nested cross-validation to classify the participants into two groups with low and high ECog scores. The input variables were the 42 acoustic features extracted from the voice responses to each question, and three demographic variables of the participant's age, sex, and years of education (45 variables in total). Because we considered the responses to each question as independent samples, the number of samples was 259 for cognitive tasks (54 participants × 5 questions – 11 missing trials) and 263 for daily life questions (54 participants × 5 questions – 7 missing trials). For reference, we also built a model using the seven neuropsychological test scores (MMSE, Frontal Assessment Battery, Clock Drawing Test, Logical Memory IA and IIA, and Trail Making Test parts A and B) in

TABLE II. MODEL PERFORMANCE FOR CLASSIFYING TWO GROUPS WITH HIGH AND LOW ECOG SCORES.

|  | ACC | SEN | SPE | F1 |
|---|---|---|---|---|
| **Neuropsychological tests (reference model)** | | | | |
| Logistic regression | 61.1 | 87.5 | 22.7 | 72.7 |
| XGBoost | 63.0 | 78.1 | 40.9 | 71.4 |
| LightGBM | 63.0 | 81.3 | 36.4 | 72.2 |
| Support vector machine | 64.8 | 68.8 | 59.1 | 69.8 |
| k-nearest neighbors | **68.5** | **68.8** | **68.2** | **72.1** |
| **Acoustic features derived from responses to the daily life questions** | | | | |
| Logistic regression | 59.3 | 71.9 | 41.0 | 67.6 |
| Support vector machine | 59.3 | 81.3 | 27.3 | 70.3 |
| k-nearest neighbors | 64.8 | 81.3 | 41.0 | 73.2 |
| XGBoost | 64.8 | 93.8 | 22.7 | 75.9 |
| LightGBM | **70.4** | **81.3** | **54.5** | **76.5** |
| **Acoustic features derived from responses to the cognitive tasks** | | | | |
| Logistic regression | 57.4 | 78.1 | 27.3 | 68.5 |
| k-nearest neighbors | 63.0 | 78.1 | 41.0 | 71.4 |
| Support vector machine | 64.8 | 84.4 | 36.4 | 74.0 |
| XGBoost | 68.5 | 81.3 | 50.0 | 75.4 |
| LightGBM | **77.8** | **87.5** | **63.6** | **82.4** |

Abbreviations: ACC, accuracy (%); SEN, sensitivity (%); SPE, specificity (%); F1 score (%)

addition to the three demographic variables. To reduce overfitting of the models using acoustic features, we applied Boruta [69] for feature selection during each fold of cross-validation. The algorithms for binary classification were k-nearest neighbors [70], logistic regression [71], support vector machine [72], XGBoost [73], and LightGBM [74]. The models were evaluated via the accuracy, sensitivity, specificity, and f1-measure. The algorithms were implemented in Python (v3.8) with the XGBoost (v1.4.2), LightGBM (v3.2.1), and scikit-learn (v1.1.3) libraries. Appendix Table I lists the hyperparameter ranges for these algorithms.

To avoid overly optimistic results, the model performance was evaluated with a 10×3 nested cross-validation procedure (Fig. 1C). The test set (one fold of data) was kept untouched and only used for performance evaluation, while hyperparameter tuning for both the feature selection and binary classifiers was performed within an inner-loop cross-validation on the training set (nine folds of data). Specifically, in the outer loop, the dataset was randomly split into training (9/10) and test (1/10) partitions, and each test partition was later used as an independent test set. In the inner loop, each training partition was further split into inner training and test folds through another three-fold cross-validation procedure to tune the hyperparameters for the feature selection and binary classification algorithms. The hyperparameter tuning was done with a Bayesian optimization algorithm, called a tree-structured Parzen estimator [75], which was implemented in the Optuna (v3.0.3) hyperparameter optimization framework library [76]. We used subject-wise, stratified sampling in the cross-validation procedure such that each fold contained approximately the same proportion of the two groups.

To compare the acoustic features that contributed to the models between the two different settings (cognitive tasks and daily conversation), we investigated the acoustic features that were robustly selected by the feature selection procedure across different training sets. Specifically, we ranked the acoustic features in accordance with the selection frequency and identified features with repeat occurrences over 50%, while aiming to exclude non-robust features. We then compared the ranked features between the two settings. We also adapted SHapley Additive exPlanations (SHAP), implemented in Python (v3.9) with the SHAP library (v0.40.0), to evaluate the importance of features in terms of their impact on the model output [77].

*F. Statistical analysis*

To corroborate the machine learning analysis, we also performed statistical analysis. Specifically, we investigated the associations of the acoustic features with the ECog scores via Spearman's rank correlation coefficient. We also used another effect size, partial eta-squared, in one-way analysis of covariance to compare the acoustic features between the groups with high and low ECog scores. We adjusted both analyses by using the age, sex, and years of education as covariates. For multiple testing of the acoustic features, the Benjamini-Hochberg correction was applied. To investigate the difference and agreement of the statistical analysis results for the different voice types, we conducted paired *t*-tests and Spearman's rank correlation tests, respectively, on the resultant correlation coefficients or effect sizes between the cognitive task and daily life question conditions. All the statistical analyses were performed in Python (v3.8) by using the Pingouin library (v0.5.0) with an alpha value of 0.05 ($p < 0.05$, two-sided).

III. RESULTS

We evaluated the model performance in discriminating the ECog low and high groups through a nested cross-validation procedure. The best model using voice responses to the cognitive tasks achieved an accuracy of 77.8% (sensitivity: 87.5%, specificity: 63.6%, F1 score: 82.4%; Table II), while the best model using responses to the daily life questions achieved an accuracy of 70.4% (sensitivity: 81.3%, specificity: 54.5%, F1 score: 76.5%; Table II). Both accuracy values were higher than the reference accuracy of 68.5% (sensitivity: 68.8%, specificity: 68.2%, F1 score: 72.1%; Table II) that was calculated by using neuropsychological test scores.

For an exploratory analysis to identify acoustic features that were discriminative across different types of voice data, we compared the selected features and their SHAP values between the models using responses to the cognitive tasks and daily life questions. In addition to the variables of age and years of education, 8 and 12 acoustic features derived from responses to the cognitive tasks and daily life questions, respectively, were selected with repeat occurrences over 50%. Interestingly, six of the eight acoustic features selected in the model using responses to the cognitive tasks were also included in the model using responses to the daily life questions (five MFCC-based features and F2). Furthermore, these features' total SHAP value constituted the majority in both models (82.1% for cognitive tasks, 52.6% for daily life questions; Fig. 2). This indicates that these acoustic features

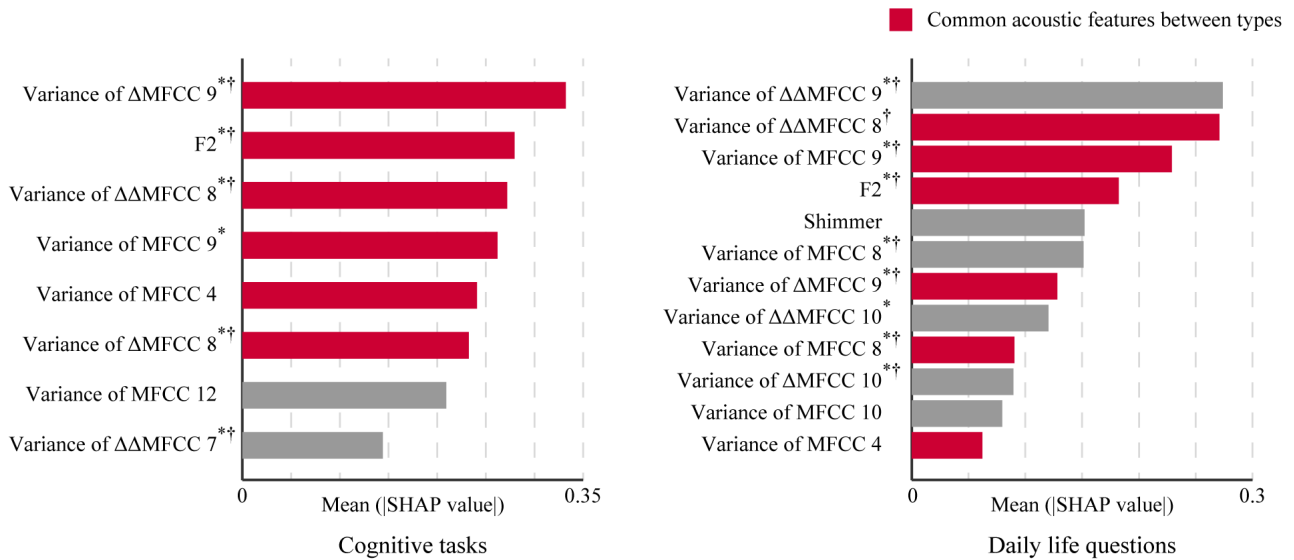

Fig. 2. Feature importance and their impact on the machine-learning model outputs. Features are selected with repeated occurrences over 50% across different training sets. The features are sorted by their mean absolute SHAP values over all samples. Red represents the features which are common across the voice responses to cognitive tasks and daily life questions. Features with an asterisk (*) represent the features with statistically significant differences between the two groups with high and low ECog scores. Features with a tagger (†) represent the features statistically correlated with the ECog scores.

can be robust markers for deficits in everyday functioning across different types of voice data.

To corroborate these results, we compared the association strength of each acoustic feature with the ECog scores between the voice responses to the cognitive tasks and daily life questions. The association was compared in terms of the correlation coefficient with the ECog scores and the effect size between the two groups with high and low ECog scores. These analyses were conducted after controlling for the age, sex, and years of education as potential confounding factors.

The analysis result showed that acoustic features derived from responses to the cognitive tasks had higher correlations with the ECog scores than those derived from responses to the daily life questions (Spearman's correlation $\rho = 0.12\pm0.07$ vs. $0.08\pm0.06$; paired $t$-test, $p < 0.001$; Fig. 3A); furthermore, the effect sizes between the groups with high and low ECog scores were also larger for acoustic features derived from responses to the cognitive tasks ($0.03\pm0.02$) as compared to those derived from responses to the daily life questions ($0.01\pm0.02$; paired $t$-test, $p < 0.001$; Fig. 4). This result aligns with an important observation regarding the model performance, where the performance with voice responses to the cognitive tasks was better than that with responses to the daily life questions for the majority of the classification algorithms used here.

From another perspective, both the correlation coefficients and effect sizes of the acoustic features were significantly correlated between the responses to the cognitive tasks and daily life questions (Spearman's correlation $\rho = 0.75$ and $0.64$; $p < .001$; Fig. 3B). This result indicates that when acoustic features extracted from voice data during cognitive tasks are discriminative for deficits in everyday functioning, these features can also be discriminative even when they are extracted from voice responses to daily life questions. In fact, among the six acoustic features selected in both models, four showed statistically significant correlations with the ECog scale (Spearman's correlation $\rho$: $0.13 < |\rho| < 0.27$; Benjamini-Hochberg adjusted $p < 0.05$; Fig. 2), as well as statistically significant differences between the high- and low-score groups (one-way analysis of covariance $\eta_p^2$: $0.02 < \eta_p^2 < 0.09$; Benjamini-Hochberg adjusted $p < 0.05$; Figs. 2 and 4), for the responses to both the cognitive tasks and daily life questions. Together, the results also indicate that common acoustic features from different types of voice data can be used for detecting deficits in everyday functioning.

IV. DISCUSSION

In this study, we collected voice responses to cognitive tasks and daily life questions with our smartwatch application and investigated the associations of acoustic features with everyday functioning in older adults as assessed by the ECog scale. We obtained two main findings. First, machine learning analysis showed that the smartwatch-derived acoustic features could differentiate individuals with high and low ECog scores with higher accuracy than the neuropsychological tests for screening cognitive impairment. Second, we found common acoustic features that contributed to both the model using responses to the cognitive tasks and the model using responses to the daily life questions. This result was further corroborated by statistical analysis of each acoustic feature, which showed statistically significant associations with the ECog scores even after excluding confounding effects, as well as strong correlations of these associations between the two different settings involving the cognitive tasks and daily life questions. These results indicate that common acoustic features can be robustly used for detecting deficits in everyday functioning from different types of voice data.

Here, we highlight the feasibility and benefits of using a smartwatch-based application to automatically detect deficits in everyday functioning. The current standard measures in both research and clinical practice are based on subjective ratings and have raised concerns about the effects of mental states such as depressive mood, anxiety, or stress [25,26].

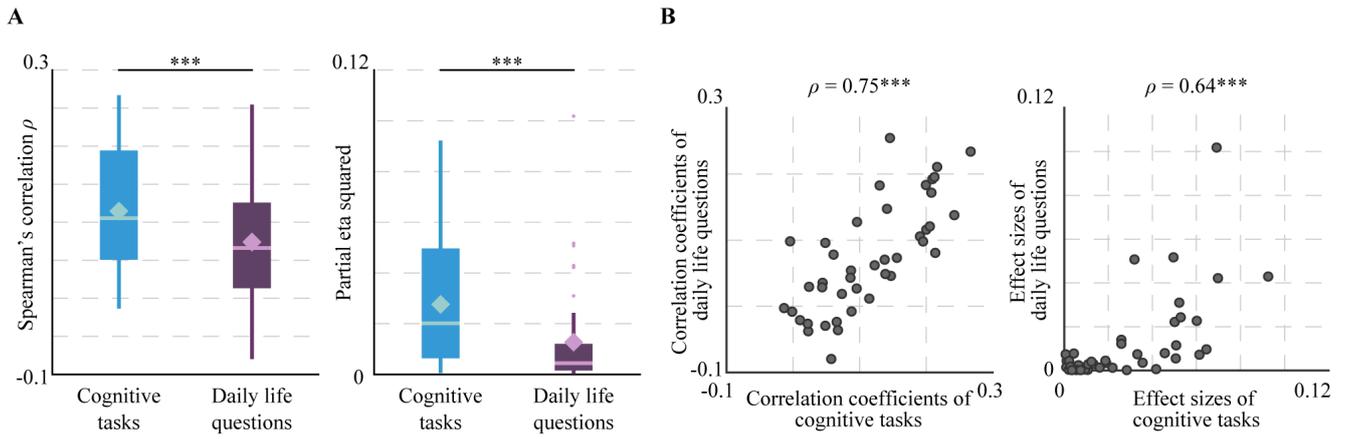

Fig. 3. Comparison of association strength of each acoustic features with the ECog scores between voice responses to the cognitive tasks and the daily life questions. **A.** Differences in the correlation coefficients with ECog scores and effect sizes between the two groups with high and low ECog scores. Boxes indicate the 25th (Q1) and 75th (Q3) percentiles. Whiskers indicate the upper and lower adjacent values that are most extreme within Q3+1.5 (Q3–Q1) and Q1–1.5 (Q3–Q1), respectively. The line and diamond in each box represent the median and mean, respectively. Dots outside of the box represent outliers. Horizontal bars indicate significant differences (paired *t*-test: ***$p < 0.001$). **B.** Correlations of the correlation coefficients with ECog scores and effect sizes between the two groups with high and low ECog scores. (Spearman's correlation $\rho$: ***$p < 0.001$).

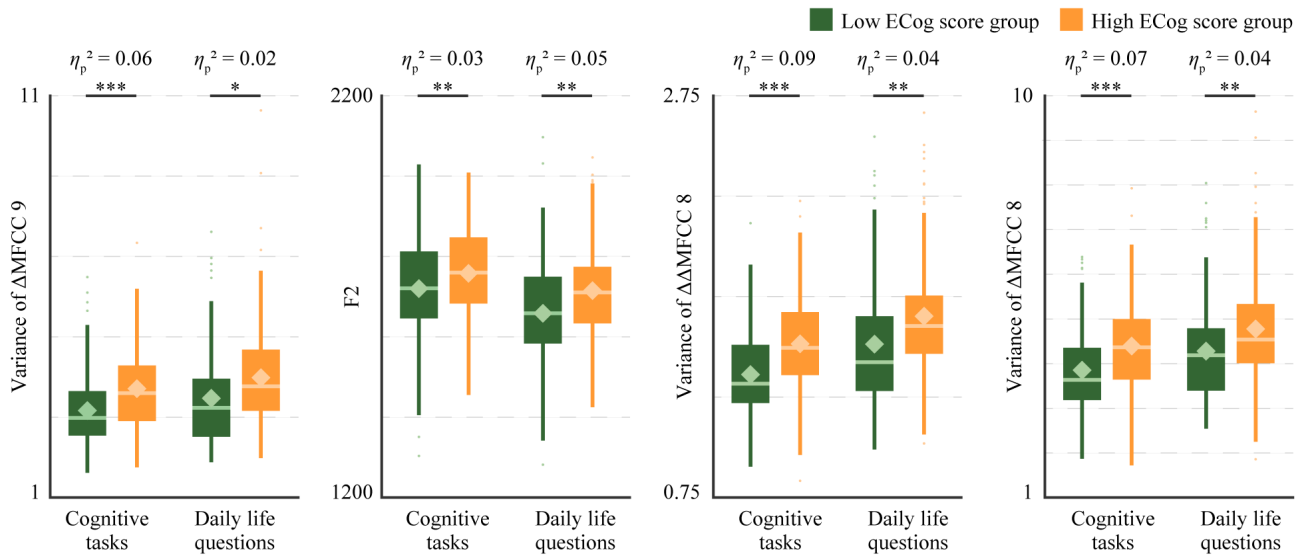

Fig. 4. Acoustic features with statistically significant correlations with the ECog scale as well as statistically significant differences between the two groups with high and low ECog scores. Differences in the correlation coefficients with ECog scores and effect sizes between the two groups with high and low ECog scores. Boxes indicate the 25th (Q1) and 75th (Q3) percentiles. Whiskers indicate the upper and lower adjacent values that are most extreme within Q3+1.5 (Q3–Q1) and Q1–1.5 (Q3–Q1), respectively. The line and diamond in each box represent the median and mean, respectively. Dots outside of the box represent outliers. Horizontal bars indicate significant differences (one-way analyses of covariance with Benjamini-Hochberg correction: *$p < 0.05$, **$p < 0.01$, ***$p < 0.001$).

Thus, the proposed objective method may offer a complementary (or possibly alternative) approach to help improve the detection accuracy for subtle changes in everyday functioning, as well as the prognostic accuracy for related diseases. In particular, assessments of everyday cognition and daily function are expected to be applied for detecting the preclinical stage of Alzheimer's disease [12,13], because clinical trials on disease-modifying drugs suggest the importance of early intervention [78,79]. Such assessments are also expected to be applied for early diagnostic differentiation of dementia subtypes, including Alzheimer's disease, dementia with Lewy bodies, and frontotemporal dementia, because they require different treatment and disease management [80–82]. From both perspectives, discriminative differences in speech and language characteristics from the cognitively normal case have been suggested in the preclinical stage of Alzheimer's disease [37] and in dementia subtypes [83]. Hence, the combination of voice-based objective measures with current subjective measures of everyday functioning may help improve the accuracy for such applications. From another viewpoint, smartwatch-based assessment using voice data from daily conversation may enable passive, continuous monitoring for deficits in everyday functioning. In fact, the use of such data collected passively with mobile devices has been explored for various types of healthcare applications, including early detection of cognitive impairments and monitoring of patients' symptoms [84,85]. Although further studies including in-situ evaluation are required, our results may help future efforts toward developing less burdensome applications that use speech data for timely detection of deficits in everyday functioning.

Our results showed that both cognitive tasks and daily life questions can elicit similar changes in acoustic features in accordance with the degree of functional limitation, and that common acoustic features can be used for detecting deficits in everyday functioning. Certain acoustic features have been reported as significant indicators for patients with cognitive impairments (e.g., Alzheimer's disease and mild cognitive impairment) across studies investigating different types of voice data collected by such means as picture description tasks, verbal fluency tasks, and open-ended interviews about daily life [49]. Although direct comparisons of acoustic changes due to cognitive impairments in different types of voice data remain limited, a few studies have suggested common acoustic features that may be robustly discriminative across different types of voice data [49,86]. Our results align with those studies and newly suggest common acoustic features that can be used for detecting deficits in cognition-relevant everyday functioning across different types of voice data, including responses to cognitive tasks and daily conversation.

We also found that both the model performance and the discriminative power of each acoustic feature were higher for voice responses to cognitive tasks than for responses to daily life questions. This indicates that acoustic features extracted from responses to cognitive tasks could show a more discernible difference in accordance with the degree of functional limitation due to cognitive impairment, as compared to features from responses to daily life questions. This finding is consistent with prior work comparing the use of cognitive tasks and daily life questions for detecting patients with mild cognitive impairment [49]. These results may be explained by the difference in cognitive load: it may be larger in cognitive tasks than in daily life questions, and this larger cognitive load may elicit larger discernible differences in acoustic features according to the degree of functional limitation due to cognitive impairment. This notion further implies that the use of acoustic features extracted from subsets of data instead of complete voice data, such as subsets involving a heavy cognitive load, may improve the accuracy for detecting functional limitations, especially when daily conversation is mixed with cognitive loads at various levels. From this perspective, a model with a self-attention mechanism [87] (e.g., a transformer), which would enable the model to focus on the most relevant voice segments, may help enable reliable detection of cognitive deficits in everyday functioning.

Our findings were derived from speech data; thus, applicability across languages is of critical interest. To this end, we expect that multiple languages could share some acoustic features that reflect speech deficits due to cognitive impairments, and thus models developed in one language would be capable of transferring to different languages. In support of this notion, so far, several unilingual studies have reported the usefulness of acoustic features in detecting cognitive impairments in their languages (e.g., Spanish [63], Turkish [88], English [35], Japanese [60], Swedish [89]), with some common trends in changes in acoustic features. Furthermore, some studies have demonstrated cross-lingual applicability of acoustic analysis models for detecting cognitive impairments, between two languages. For example, Bertini, *et al.* [90] showed that their model using spectrum-based features achieved good accuracies between English and Italian datasets in classifying patients with Alzheimer's disease from controls. Also, Tamm, *et al.* [91] showed that their deep learning-based model using acoustic features could successfully transfer from English to Greek. These results from previous studies suggest the existence of common acoustic changes due to cognitive impairments across languages, which would support the potential cross-language applicability of our acoustic analysis model for cognition-relevant everyday functioning.

This study had several limitations. First, we collected the speech data in a lab setting, and the controlled setting might have affected how the participants responded to questions. Our findings require further validation in situ. Second, to our knowledge this is the first empirical study investigating the association of voice features with everyday functioning in older adults; however, the limited sample size might have affected our results' generalizability.

In summary, this pilot study has provided initial empirical results suggesting the feasibility of using a smartwatch-based application to detect deficits in everyday functioning by exploiting common acoustic markers derived from different types of voice data, which are not limited to data collected during cognitive task performance. Our findings may help promote future efforts toward the early detection of functional limitations in daily life and related neurodegenerative diseases of aging, particularly Alzheimer's disease.

REFERENCES

[1] S. T. Farias, D. Mungas, B. R. Reed, D. Cahn-Weiner, W. Jagust et al., "The measurement of everyday cognition (ECog): scale development and psychometric properties." Neuropsychology, vol. 22, no. 4, p. 531, 2008.

[2] C. M. Giebel, C. Sutcliffe, M. Stolt, S. Karlsson, A. Renom-Guiteras et al., "Deterioration of basic activities of daily living and their impact on quality of life across different cognitive stages of dementia: a european study," International psychogeriatrics, vol. 26, no. 8, pp. 1283–1293, 2014.

[3] C. M. Giebel, C. Sutcliffe, and D. Challis, "Activities of daily living and quality of life across different stages of dementia: a UK study," Aging & Mental health, vol. 19, no. 1, pp. 63–71, 2015.

[4] B. A. Springate and G. Tremont, "Dimensions of caregiver burden in dementia: Impact of demographic, mood, and care recipient variables," The Amercan Journal of Geriatric Psychiatry, vol. 22, no. 3, pp. 294–300, 2014

[5] S. T. Farias, D. Mungas, B. R. Reed, D. Harvey, D. Cahn-Weiner et al., "MCI is asociated with deficits in everyday functioning," Alzheimer disease and associated disorders, vol. 20, no. 4, p. 217, 2006.

[6] S. T. Farias, E. Chou, D. J. Harvey, D. Mungas, B. Reed et al., "Longitudinal trajectories of everyday function by diagnostic status." Psycholoy and aging, vol. 28, no. 4, p. 1070, 2013.

[7] S. Reppermund, H. Brodaty, J. Crawford, N. Kochan, B. Draper et al., "Impairment in instrumental activities of daily living with high cognitive demand is an early marker of mild cognitive impairment: the sydney memory and ageing study," Psychological medicine, vol. 43, no. 11, pp. 2437–2445, 2013.

[8] P. Scheltens, B. De Strooper, M. Kivipelto, H. Holstege, G. Chételat et al., "Alzheimer's disease," The Lancet, vol. 397, no. 10284, pp. 1577– 1590, 2021.

[9] R. A. Sperling, P. S. Aisen, L. A. Beckett, D. A. Bennett, S. Craft et al., "Toward defining the preclinical stages of Alzheimer's disease: Recommendations from the National Institute on Aging-Alzheimer's Association workgroups on diagnostic guidelines for Alzheimer's disease," Alzheimer's & dementia, vol. 7, no. 3, pp. 280–292, 2011.

[10] D. A. Gold, "An examination of instrumental activities of daily living assessment in older adults and mild cognitive impairment," Journal of clinical and experimental neuropsychology, vol. 34, no. 1, pp. 11–34, 2012.

[11] K. Jekel, M. Damian, C. Wattmo, L. Hausner, R. Bullock et al., "Mild cognitive impairment and deficits in instrumental activities of daily living: a systematic review," Alzheimer's research & therapy, vol. 7, no. 1, pp. 1–20, 2015.

[12] G. A. Marshall, S. A. Sikkes, R. E. Amariglio, J. R. Gatchel, D. M. Rentz et al., "Instrumental activities of daily living, amyloid, and cognition in cognitively normal older adults screening for the a4 study," Alzheimer's & Dementia: Diagnosis, Assessment & Disease Monitoring, vol. 12, no. 1, p. e12118, 2020.

[13] M. Lilamand, M. Cesari, C. Cantet, P. Payoux, S. Andrieu et al., "Relationship between brain amyloid deposition and instrumental activities of daily living in older adults: a longitudinal study from the multidomain alzheimer prevention trial," Journal of the American Geriatrics Society, vol. 66, no. 10, pp. 1940–1947, 2018.

[14] S. T. Farias, T. Giovannetti, B. R. Payne, M. Marsiske, G. W. Rebok et al., "Self-perceived difficulties in everyday function precede cognitive decline among older adults in the active study," Journal of the International Neuropsychological Society, vol. 24, no. 1, pp. 104–112, 2018.

[15] J. Verghese, P. De Sanctis, and E. Ayers, "Everyday function profiles in prodromal stages of mci: Prospective cohort study," Alzheimer's & Dementia, vol. 19, no. 2, pp. 498–506, 2023.

[16] S. T. Farias, D. Mungas, B. R. Reed, D. Harvey, and C. DeCarli, "Progression of mild cognitive impairment to dementia in clinic-vs community-based cohorts," Archives of neurology, vol. 66, no. 9, pp. 1151–1157, 2009.

[17] S. T. Farias, K. Lau, D. Harvey, K. G. Denny, C. Barba et al., "Early functional limitations in cognitively normal older adults predict diagnostic conversion to mild cognitive impairment," Journal of the American Geriatrics Society, vol. 65, no. 6, pp. 1152–1158, 2017.

[18] J. J. Gomar, P. D. Harvey, M. T. Bobes-Bascaran, P. Davies, and T. E. Goldberg, "Development and cross-validation of the upsa short form for the performance-based functional assessment of patients with mild cognitive impairment and alzheimer disease," The American journal of geriatric psychiatry, vol. 19, no. 11, pp. 915–922, 2011.

[19] G. A. Marshall, M. Dekhtyar, J. M. Bruno, K. Jethwani, R. E. Amariglio et al., "The harvard automated phone task: new performance-based activities of daily living tests for early Alzheimer's disease," The journal of prevention of Alzheimer's disease, vol. 2, no. 4, p. 242, 2015.

[20] M. J. Sliwinski, J. A. Mogle, J. Hyun, E. Munoz, J. M. Smyth et al., "Reliability and validity of ambulatory cognitive assessments," Assessment, vol. 25, no. 1, pp. 14–30, 2018.

[21] S. A. Sikkes, E. S. de Lange-de Klerk, Y. A. Pijnenburg, F. Gillissen, R. Romkes et al., " A new informant-based questionnaire for instrumental activities of daily living in dementia," Alzheimer's & Dementia, vol. 8, no. 6, pp. 536‑543, 2012.

[22] A. D. Rueda, K. M. Lau, N. Saito, D. Harvey, S. L. Risacher et al., "Self-rated and informant-rated everyday function in comparison to objective markers of Alzheimer's disease," Alzheimer's & Dementia, vol. 11, no. 9, pp. 1080–1089, 2015.

[23] M. A. Dubbelman, J. Sanchez, A. P. Schultz, D. M. Rentz, R. E. Amariglio et al., "Everyday functioning and entorhinal and inferior temporal tau burden in cognitively normal older adults," The Journal of Prevention of Alzheimer's Disease, vol. 9, no. 4, pp. 801–808, 2022.

[24] A. Martyr, S. M. Nelis, and L. Clare, "Predictors of perceived functional ability in early-stage dementia: self-ratings, informant ratings and discrepancy scores," International journal of geriatric psychiatry, vol. 29, no. 8, pp. 852–862, 2014.

[25] S. Weintraub, M. C. Carrillo, S. T. Farias, T. E. Goldberg, J. A. Hendrix et al., "Measuring cognition and function in the preclinical stage of Alzheimer's disease," Alzheimer's & Dementia: Translational Research & Clinical Interventions, vol. 4, pp. 64–75, 2018.

[26] L. C. Kourtis, O. B. Regele, J. M. Wright, and G. B. Jones, "Digital biomarkers for alzheimer's disease: the mobile/wearable devices opportunity," NPJ digital medicine, vol. 2, no. 1, p. 9, 2019.

[27] L. Masanneck, P. Gieseler, W. J. Gordon, S. G. Meuth, and A. D. Stern, "Evidence from clinicaltrials.gov on the growth of digital health technologies in neurology trials," npj Digital Medicine, vol. 6, no. 1, p. 23, 2023.

[28] L. L. Law et al., "Moderate intensity physical activity associates with CSF biomarkers in a cohort at risk for Alzheimer's disease," Alzheimer's & Dementia: Diagnosis, Assessment & Disease Monitoring, vol. 10, pp. 188–195, 2018.

[29] J. F. Thayer, F. Åhs, M. Fredrikson, J. J. Sollers, and T. D. Wager, "A meta-analysis of heart rate variability and neuroimaging studies: Implications for heart rate variability as a marker of stress and health," Neuroscience & Biobehavioral Reviews, vol. 36, no. 2, pp. 747–756, 2012.

[30] D. Liaqat, R. Wu, A. Gershon, H. Alshaer, F. Rudzicz, and E. De Lara, "Challenges with real-world smartwatch based audio monitoring," in Proceedings of the 4th ACM Workshop on Wearable Systems and Applications, Munich Germany: ACM, 2018, pp. 54–59.

[31] M. Nicholas, L. K. Obler, M. L. Albert, and N. Helm-Estabrooks, "Empty speech in Alzheimer's disease and fluent aphasia," Journal of Speech, Language, and Hearing Research, vol. 28, no. 3, pp. 405–410, 1985.

[32] C. K. Tomoeda and K. A. Bayles, "Longitudinal effects of Alzheimer disease on discourse production." Alzheimer Disease and Associated Disorders, 1993.

[33] S. Ahmed, A.-M. F. Haigh, C. A. de Jager, and P. Garrard, "Connected speech as a marker of disease progression in autopsy-proven Alzheimer's disease," Brain, vol. 136, no. 12, pp. 3727–3737, 2013.

[34] E. Eyigoz, S. Mathur, M. Santamaria, G. Cecchi, and M. Naylor, "Linguistic markers predict onset of Alzheimer's disease," EClinicalMedicine, vol. 28, p. 100583, 2020.

[35] K. C. Fraser, J. A. Meltzer, and F. Rudzicz, "Linguistic features identify Alzheimer's disease in narrative speech." J Alzheimers Dis, vol. 49, no. 2, pp. 407–422, 2016.

[36] L. Hernández-Domínguez, S. Ratté, G. Sierra-Martínez, and A. Roche-Bergua, "Computer-based evaluation of Alzheimer's disease and mild cognitive impairment patients during a picture description task," Alzheimer's & Dementia: Diagnosis, Assessment & Disease Monitoring, vol. 10, pp. 260–268, 2018.

[37] F. Cuetos, J. C. Arango-Lasprilla, C. Uribe, C. Valencia, and F. Lopera, "Linguistic changes in verbal expression: a preclinical marker of Alzheimer's disease," Journal of the International Neuropsychological Society, vol. 13, no. 3, pp. 433–439, 2007.

[38] Y. Yamada, K. Shinkawa, and K. Shimmei, "Atypical repetition in daily conversation on different days for detecting Alzheimer disease: evaluation of phone-call data from a regular monitoring service," JMIR mental health, vol. 7, no. 1, p. e16790, 2020.

[39] A. Balagopalan, B. Eyre, J. Robin, F. Rudzicz, and J. Novikova, "Comparing pre-trained and feature-based models for prediction of Alzheimer's disease based on speech," Frontiers in aging neuroscience, vol. 13, p. 635945, 2021.

[40] Y. Momota, K.-c. Liang, T. Horigome, M. Kitazawa, Y. Eguchi et al., "Language patterns in japanese patients with alzheimer disease: A machine learning approach," Psychiatry and Clinical Neurosciences, 2022.

[41] A. A. Wisler, A. R. Fletcher, and M. J. McAuliffe, "Predicting montreal cognitive assessment scores from measures of speech and language," Journal of Speech, Language, and Hearing Research, vol. 63, no. 6, pp. 1752–1761, 2020.

[42] S. Ash, P. Moore, L. Vesely, and M. Grossman, "The decline of narrative discourse in Alzheimer's disease," Brain and Language, vol. 1, no. 103, pp. 181–182, 2007.

[43] C. Drummond, G. Coutinho, R. P. Fonseca, N. Assunção, A. Teldeschi et al., "Deficits in narrative discourse elicited by visual stimuli are already present in patients with mild cognitive impairment," Frontiers in aging neuroscience, vol. 7, p. 96, 2015.

[44] C. Diaz-Asper, C. Chandler, R. S. Turner, B. Reynolds, and B. Elvevåg, "Increasing access to cognitive screening in the elderly: Applying natural language processing methods to speech collected over the telephone," Cortex, vol. 156, pp. 26–38, 2022.

[45] M. Kobayashi, A. Kosugi, H. Takagi, M. Nemoto, K. Nemoto et al., "Effects of age-related cognitive decline on elderly user interactions with voice-based dialogue systems," in IFIP Conference on Human-Computer Interaction, 2019, pp. 53–74.

[46] X. Liang, J. A. Batsis, Y. Zhu, T. M. Driesse, R. M. Roth et al., "Evaluating voice-assistant commands for dementia detection," Computer Speech & Language, vol. 72, p. 101297, 2022.

[47] K. Yoshii, D. Kimura, A. Kosugi, K. Shinkawa, T. Takase et al., "Screening of mild cognitive impairment through conversations with humanoid robots: Exploratory pilot study," JMIR Formative Research, vol. 7, no. 1, p. e42792, 2023

[48] S. T. Farias, D. Mungas, D. J. Harvey, A. Simmons, B. R. Reed et al., "The measurement of everyday cognition: development and validation

of a short form of the everyday cognition scales," Alzheimer's & Dementia, vol. 7, no. 6, pp. 593–601, 2011.

[49] Y. Yamada, K. Shinkawa, M. Kobayashi, M. Nishimura, M. Nemoto et al., "Tablet-based automatic assessment for early detection of Alzheimer's disease using speech responses to daily life questions," Frontiers in Digital Health, vol. 3, p. 30, 2021.

[50] H. Goodglass, E. Kaplan, and S. Weintraub, BDAE: The Boston Diagnostic Aphasia Examination. Lippincott Williams & Wilkins Philadelphia, PA, 2001.

[51] A. König, A. Satt, A. Sorin, R. Hoory, A. Derreumaux et al., "Use of speech analyses within a mobile application for the assessment of cognitive impairment in elderly people," Current Alzheimer Research, vol. 15, no. 2, pp. 120–129, 2018.

[52] S. Nasreen, J. Hough, M. Purver et al., "Detecting Alzheimer's disease using interactional and acoustic features from spontaneous speech." Interspeech, 2021.

[53] P. Mermelstein, "Distance measures for speech recognition, psychological and instrumental." Pattern recognition and artificial intelligence, vol. 116, pp. 374–388, 1976.

[54] N. Cummins, S. Scherer, J. Krajewski, S. Schnieder, J. Epps et al., "A review of depression and suicide risk assessment using speech analysis." Speech Commun, vol. 71, pp. 10–49, 2015.

[55] Y. Yamada, K. Shinkawa, M. Nemoto, K. Nemoto, and T. Arai, "A mobile application using automatic speech analysis for classifying Alzheimer's disease and mild cognitive impairment," Computer Speech & Language, vol. 81, p. 101514, 2023.

[56] W. H. Press and S. A. Teukolsky, "Savitzky-golay smoothing filters," Computers in Physics, vol. 4, no. 6, pp. 669-672, 1990.

[57] N. Roy, S. L. Nissen, C. Dromey, and S. Sapir, "Articulatory changes in muscle tension dysphonia: evidence of vowel space expansion following manual circumlaryngeal therapy," Journal of communication disorders, vol. 42, no. 2, pp. 124–135, 2009.

[58] Y. Yamada, K. Shinkawa, M. Nemoto, and T. Arai, "Automatic Assessment of Loneliness in Older Adults Using Speech Analysis on Responses to Daily Life Questions," Frontiers in Psychiatry, vol. 12, p. 2294, 2021.

[59] J. P. Teixeira, C. Oliveira, and C. Lopes, "Vocal acoustic analysis–jitter, shimmer and hnr parameters," Procedia Technology, vol. 9, pp. 1112–1122, 2013.

[60] A. O. Hall, K. Shinkawa, A. Kosugi, T. Takase, M. Kobayashi et al., "Using tablet-based assessment to characterize speech for individuals with dementia and mild cognitive impairment: preliminary results," AMIA Summits on Translational Science Proceedings, vol. 2019, p. 34, 2019.

[61] M. Ogawa, G. Oyama, K. Morito, M. Kobayashi, Y. Yamada et al., "Can ai make people happy? the effect of AI-based chatbot on smile and speech in parkinson's disease," Parkinsonism & Related Disorders, vol. 99, pp. 43–46, 2022.

[62] J. Rusz, R. Cmejla, H. Ruzickova, and E. Ruzicka, "Quantitative acoustic measurements for characterization of speech and voice disorders in early untreated Parkinson's disease," The journal of the Acoustical Society of America, vol. 129, no. 1, pp. 350–367, 2011.

[63] J. J. G. Meilán, F. Martínez-Sánchez, J. Carro, D. E. López, L. Millian-Morell et al., "Speech in Alzheimer's disease: Can temporal and acoustic parameters discriminate dementia?" Dement Geriatr Cogn Disord, vol. 37, no. 5-6, pp. 327–334, 2014.

[64] N. D. Pah, M. A. Motin, and D. K. Kumar, "Phonemes based detection of parkinson's disease for telehealth applications," Scientific Reports, vol. 12, no. 1, p. 9687, 2022.

[65] A. König, N. Linz, R. Zeghari, X. Klinge, J. Tröger et al., "Detecting apathy in older adults with cognitive disorders using automatic speech analysis," Journal of Alzheimer's Disease, vol. 69, no. 4, pp. 1183–1193, 2019.

[66] T. F. Quatieri and N. Malyska, "Vocal-source biomarkers for depression: A link to psychomotor activity," in Thirteenth annual conference of the international speech communication association, 2012.

[67] B. McFee, C. Raffel, D. Liang, D. P. Ellis, M. McVicar et al., "librosa: Audio and music signal analysis in python." in Proceedings of the 14th python in science conference, vol. 8. Citeseer, 2015, pp. 18–25.

[68] Signal_analysis. https://brookemosby.github.io/Signal_Analysis/ [accessed 2023-2-28].

[69] M. B. Kursa and W. R. Rudnicki, "Feature selection with the boruta package," Journal of statistical software, vol. 36, pp. 1–13, 2010.

[70] J. Goldberger, G. E. Hinton, S. Roweis, and R. R. Salakhutdinov, "Neighbourhood components analysis." Adv Neural Inf Process Syst, vol. 17, pp. 513–520, 2004.

[71] D. R. Cox, "The regression analysis of binary sequences," Journal of the Royal Statistical Society: Series B (Methodological), vol. 20, no. 2, pp. 215–232, 1958.

[72] B. E. Boser, I. M. Guyon, and V. N. Vapnik, "A training algorithm for optimal margin classifiers." in Proceedings of the fifth annual workshop on Computational learning theory, 1992, pp. 144–152.

[73] T. Chen and C. Guestrin, "Xgboost: A scalable tree boosting system," in Proceedings of the 22nd acm sigkdd international conference on knowledge discovery and data mining, 2016, pp. 785–794.

[74] G. Ke, Q. Meng, T. Finley, T. Wang, W. Chen et al., "Lightgbm: A highly efficient gradient boosting decision tree," Advances in neural information processing systems, vol. 30, 2017.

[75] J. Bergstra, D. Yamins, and D. Cox, "Making a science of model search: Hyperparameter optimization in hundreds of dimensions for vision architectures," in International conference on machine learning. PMLR, 2013, pp. 115–123.

[76] T. Akiba, S. Sano, T. Yanase, T. Ohta, and M. Koyama, "Optuna: A next-generation hyperparameter optimization framework," in Proceedings of the 25th ACM SIGKDD international conference on knowledge discovery & data mining, 2019, pp. 2623–2631.

[77] S. M. Lundberg and S.-I. Lee, "A unified approach to interpreting model predictions," in Proceedings of the 31st international conference on neural information processing systems, 2017, pp. 4768–4777.

[78] S. Budd Haeberlein, P. Aisen, F. Barkhof, S. Chalkias, T. Chen et al., "Two randomized phase 3 studies of aducanumab in early alzheimer's disease," The Journal of Prevention of Alzheimer's Disease, vol. 9, no. 2, pp. 197–210, 2022.

[79] C. H. Van Dyck, C. J. Swanson, P. Aisen, R. J. Bateman, C. Chen et al., "Lecanemab in early Alzheimer's disease," New England Journal of Medicine, vol. 388, no. 1, pp. 9–21, 2023.

[80] J. M. Hamilton, D. P. Salmon, R. Raman, L. A. Hansen, E. Masliah et al., "Accounting for functional loss in Alzheimer's disease and dementia with lewy bodies: beyond cognition," Alzheimer's & Dementia, vol. 10, no. 2, pp. 171–178, 2014.

[81] D. P. Gill, R. A. Hubbard, T. D. Koepsell, M. J. Borrie, R. J. Petrella et al., "Differences in rate of functional decline across three dementia types," Alzheimer's & Dementia, vol. 9, no. 5, pp. S63–S71, 2013.

[82] N. Moheb, M. F. Mendez, S. A. Kremen, and E. Teng, "Executive dysfunction and behavioral symptoms are associated with deficits in instrumental activities of daily living in frontotemporal dementia," Dementia and geriatric cognitive disorders, vol. 43, no. 1-2, pp. 89– 99, 2017.

[83] Y. Yamada, K. Shinkawa, M. Nemoto, M. Ota, K. Nemoto et al., "Speech and language characteristics differentiate Alzheimer's disease and dementia with lewy bodies," Alzheimer's & Dementia: Diagnosis, Assessment & Disease Monitoring, vol. 14, no. 1, p. e12364, 2022.

[84] M. Faurholt-Jepsen, J. Busk, M. Frost, M. Vinberg, E. M. Christensen et al., "Voice analysis as an objective state marker in bipolar disorder," Translational psychiatry, vol. 6, no. 7, pp. e856–e856, 2016.

[85] D. Di Matteo, K. Fotinos, S. Lokuge, G. Mason, T. Sternat et al., "Automated screening for social anxiety, generalized anxiety, and depression from objective smartphone-collected data: cross-sectional study," Journal of Medical Internet Research, vol. 23, no. 8, p. e28918, 2021.

[86] N. Clarke, T. R. Barrick, and P. Garrard, "A comparison of connected speech tasks for detecting early Alzheimer's disease and mild cognitive impairment using natural language processing and machine learning," Frontiers in Computer Science, vol. 3, p. 634360, 2021.

[87] A. Vaswani, N. Shazeer, N. Parmar, J. Uszkoreit, L. Jones et al., "Attention is all you need," Advances in neural information processing systems, vol. 30, 2017.

[88] A. Khodabakhsh, F. Yesil, E. Guner, and C. Demiroglu, "Evaluation of linguistic and prosodic features for detection of Alzheimer's disease in Turkish conversational speech," EURASIP Journal on Audio, Speech, and Music Processing, vol. 2015, no. 1, p. 9, 2015.

[89] C. Themistocleous, M. Eckerström, and D. Kokkinakis, "Voice quality and speech fluency distinguish individuals with Mild Cognitive

Impairment from Healthy Controls," PLoS ONE, vol. 15, no. 7, p. e0236009, 2020.
[90] F. Bertini, D. Allevi, G. Lutero, L. Calzà, and D. Montesi, "A Cross-language Dementia Classifier: a Preliminary Study," in 2022 IEEE International Conference on Metrology for Extended Reality, Artificial Intelligence and Neural Engineering (MetroXRAINE), 2022, pp. 438–443.
[91] B. Tamm, R. Vandenberghe, and H. Van Hamme, "Cross-Lingual Transfer Learning for Alzheimer's Detection from Spontaneous Speech," in ICASSP 2023 - 2023 IEEE International Conference on Acoustics, Speech and Signal Processing (ICASSP), 2023, pp. 1–2.

APPENDIX TABLE I: Hyperparameters for machine learning models

| Model | Hyper-parameter | Range |
|---|---|---|
| **k-nearest neighbors** | n_neighbors | 1 to 20 |
| | weights | uniform, distance |
| | metric | euclidean, manhattan, minkowski |
| **Logistic regression** | penalty | elasticnet, l1, l2, none |
| | C | $1\times10^0$ to $1\times10^4$ |
| | l1_ratio | 0.1 to 0.9 |
| | solver | saga |
| **Support vector machine** | kernel | rbf, linear |
| | C | $1\times10^{-2}$ to $1\times10^2$ |
| | gamma | $1\times10^{-3}$ to $1\times10^0$ |
| **XGBoost** | booster | gbtree, gblinear, dart |
| | lambda | 1 to 4 |
| | alpha | $1\times10^{-8}$ to $1\times10^2$ |
| | subsample | 0.5 to 1.0 |
| | colsample_by_tree | 0.5 to 1.0 |
| | max_depth | 1 to 11 |
| | min_child_weight | 1 to $1\times10^2$ |
| | eta | $1\times10^{-8}$ to 1.0 |
| | gamma | $1\times10^{-8}$ to 7 |
| | grow_ploicy | depthwise, lossguide |
| | sample_type | uniform, weighted |
| | normalize_type | tree, forest |
| | rate_drop | $1\times10^{-8}$ to 1.0 |
| | skip_drop | $1\times10^{-8}$ to 1.0 |
| **LightGBM** | lambda_l1 | $1\times10^0$ to $1\times10^1$ |
| | lambda_l2 | $1\times10^{-2}$ to $1\times10^0$ |
| | num_leaves | 10 to 32 |
| | feature_fraction | 0.1 to 0.5 |
| | bagging_fraction | 0.8 to 1.0 |
| | bagging_freq | 3 to 7 |
| | min_child_samples | 1 to 16 |